\documentclass[a4paper,12pt]{article}
\usepackage{graphicx}

\title{Nonlinear evolution of coarse-grained quantum systems with generalized purity constraints}
\author{Nikola Buri\' c {\thanks {buric@ipb.ac.rs}}\\ Institute of Physics, \\PO Box 68, 11000 Belgrade, Serbia.}

\begin{document}

\maketitle

\begin{abstract}

Constrained quantum dynamics is used to propose a nonlinear dynamical equation for
 pure states of a generalized coarse-grained system. The relevant constraint
 is given either by the generalized purity or by the generalized
 invariant fluctuation, and the coarse-grained pure states correspond to
the generalized coherent i.e. generalized nonentangled states. Open system model
 of the coarse-graining is discussed. It is shown that in this model and in the
 weak coupling limit the constrained dynamical equations coincide with an equation
 for pointer states, based on Hilbert-Schmidt distance, that was previously suggested
 in the context of the decoherence theory.

\end{abstract}

\section{Introduction}

Coarse-grained description of a dynamical system is based on a separation of observables into two classes:
 the class of distinguished i.e. important observables and the class of observables that are
 considered inaccessible. A new system, whose
 state can be maximally determined by the preferred observables, is then defined. Evolution of the new, coarse-grained,
 system should be completely described in terms of the distinguished observables only.

From an operational point of view, the choice of distinguished observables is dictated in practice by what can be
 measured on the given system. For example: a) If all observables of a quantum system with the Hilbert space with
$N$ complex dimensions $H^N$ are considered experimentally accessible then every Hermitian operator represents an
 observable of the system, i.e. all observables are distinguished, and the pure states of the system are by definition
rays in the Hilbert space $H^N$; b) In the case of two spatially separated qubits $H^4=H_1^2\otimes H_2^2$ one could
 consider experimentally accessible only the local observables $\sigma^1_i\otimes 1$ and $1\otimes \sigma^2_i$. In this
 case the coarse-grained states are the product states; c) In a collection of $n$ spins $H^{2^n}=H^2\otimes H^2\otimes\dots H^2$ one
 might be able to observe only the  macroscopic magnetizations $m_i=\sum \sigma_i/n$. which are then the coarse-grained
 distinguished observables

 The set of distinguished observables as a subset of the algebra $U(N)$ is important in the definitions of notions such as 
quantum degrees of freedom and quantum integrability \cite{Chang}, generalized coherent states\cite{Perelomov,Gilmor}
and generalized entanglement\cite{Viola1,Viola2,Viola3,Klyasko}, and provides a framework to study the relations between these
 notions \cite{japrep}. In particular we shall be interested in the coarse-grained states representing the generalized non-entangled
states as introduced and studied in \cite{Viola1,Viola2,Viola3}. Our goal is to derive an evolution equation for which the set
of g-nonentangled states is invariant for arbitrary Hamiltonian, and discuss its physical interpretation.

In the next section we recapitulate the theory of generalized entanglement and generalized purity. In section 3 we treat a quantum
dynamical system on $H^N$ as a classical Hamiltonian system on $R^{2N}$, which enables us to discuss constrained quantum dynamics.
 This is used to derive an evolution equation of the states which preserves the maximal generalized purity, that is of the g-nonentangled states.
 This evolution equation is nonlinear and can generate, depending on the Hamiltonian,  chaotic dynamics of the coarse-grained system.
 Open quantum system model of the evolution of the distinguished states is discussed in section 4. In this section we show that
our constrained evolution equation coincides in the weak coupling
limit with the approximate
 evolution equation of the robust states derived in \cite{Strunz} by different means and in the context of decoherence theory.

\section{Generalized entanglement and generalized purity}

A selected set of distinguished observables is used to define the generalized notions of non-entangled and entangled states.
 The coarse-graining by the distinguished observables, understood in the traditional probabilistic sens as replacing probabilities
by conditional probabilities is crucial in this definition.

Consider a subset $g\in u(N)$ of distinguished observables. A state $\rho_{g}$ is called $g$-reduced state of the state $\rho$
 if $Tr[\rho L_l]=Tr[\rho_g L_l]$ for any $L_l\in g$. The reduced state $\rho_{g}$ is the projection of the state $\rho$ on the subspace
determined by distinguished observables. Identifying the quantum states with probabilities the standard definition of the conditional
 probability is recognized.
Pure state $\rho=|\psi><\psi|$ is generalized non-entangled if the corresponding reduced state
$\rho_{g}$ is pure $\rho_g^2=\rho_g$. Otherwise the pure state $|\psi>$ is g-entangled. In the case that the Hilbert space has the
bipartite tensor product structure and each distinguished observable act nontrivially only in one of the components, the previous definition of g-entanglement reduces to the standard definition of the bipartite entanglement for pure states.

In a large class of situations of physical interest the set of distinguished observables forms a Lie algebra.
In this case
a measure of the generalized entanglement of the pure state $|\psi>$ is provided by the generalized purity,
which is the purity of the reduced state $\rho_g$, and is given by:
\begin{equation}
 P_g(\psi)=\sum_l <\psi|L_l|\psi>^2,\qquad L_l\in g
\end{equation}
where $L_l$ form a bases of the Lie algebra $g$. The state $|\psi>$ is generalized non-entangled if $P_g(\psi)$ is maximal.
 Pure states with $P_g(\psi)$ less then maximal represent g-entangled states, i.e. the states in which the coarse grained system g
 is entangled with the environment, i.e. with the operators not in $g$.
 Obviously, whether a pure state $|\psi>\in H^N$ is generalized entangled or not depends on the
 choice of the distinguished observables. Once the distinguished observables are chosen, the question if the future orbit of a
g-nonentangled $|\psi>$ will remain in the set of g-nonentangled states depends on the evolution equation satisfied by $|\psi>$.

An equivalent measures of g-entanglement is given by the total dispersion of the algebra of distinguished observables
\begin{equation}
\Delta_g(\psi)=\sum_l <L_l^2>-<L_l>^2=\sum_l(\Delta L_l)^2.
\end{equation}
$\Delta_g(\psi)$ is minimal iff $P_g(\psi)$ is maximal. Expressions for the minimal value of
$\Delta_g(\psi)$ and the maximal value of $P_g(\psi)$  in terms of the simple roots of $g$ are known \cite {Viola4,Klyasko} and read
\begin{equation}
 \Delta_q(\psi)\geq \sum_l k_l<\alpha_l,\alpha_l>\equiv min,\qquad P_q(\psi)\leq <C^2>-min\equiv max
\end{equation}
where the highest weight vector $\lambda=\sum_l k_l\alpha_l$ in terms of simple roots $\alpha_l$ and
$C$ is the quadratic Casimir operator.

Generalized coherent states have been defined for an arbitrary semi-simple Lie algebra.
If the algebra
 of distinguished observables $g$ is semi-simple the minimum of $\Delta_q(\psi)$ and the maximum of
 $P_q(\psi)$ is achieved on
 the corresponding generalized coherent states \cite{Delburgo}.
 Thus, in this case the class of
 g-nonentangled and g-coherent states coincide.

\section{Evolution equation of the g-nonentangled pure states}

In general, reduction of the pure state $|\psi>$ results in a mixed state $\rho_g$ and the unitary Schroedinger evolution
of $|\psi>$ upon reduction becomes nonunitary, resulting in different forms (under different approximations) of
 master equations for $\rho_g(t)$. However, if the g-nonentangled pure state $|\psi>$ evolves in the subset
of g-nonentangled states the reduced state always remains pure. In order for this to occur in general the Hamiltonian
 linear evolution of $|\psi(t)>$ is not enough and a nonlinear constrain has to be added to ensure the preservation of
the g-purity $P_g(\psi(t))$. In order to formulate such constrained evolution we shall use the classical Hamiltonian formulation
of the quantum evolution.

It is well known (please see \cite{Hughston_geo} or \cite{JaAnnPhys} and references therein) that the evolution of a
 quantum pure state in $H^N$ as given by
 the Schroedinger equation
 can be equivalently described by a Hamiltonian dynamical system on $R^{2N}$ with the evolution equations in the Hamiltonian form:
\begin{equation}
 \dot x^i=\omega^{ij}\nabla_j {\cal H},
\end{equation}
where $x^i=q^i=(c_i^{*}+c_i)/\sqrt 2,\> i=1,2\dots N; \>x^i=p^i={\sqrt -1}(c_i^{*}-c_i)/\sqrt 2,\> i=N+1,2\dots 2N$ is the vector
 of coordinates $q_i$ and momenta $p_i$, and $c_i$ are complex expansion coefficients of the pure state $|\psi>$ in some basis.
The Hamilton's function ${\cal H}(x)$ is given by the quantum expectation of the Hamiltonian $H$ in the state
$|\psi>$: ${\cal H}=<\psi|H|\psi>$,
 and the inverse of the symplectic form $\omega^{ij}$ is given by the imaginary part of the scalar product in $H^N$. In the canonical
coordinates $x_i$ the symplectic form  $\omega_{ij}$ assumes the standard form
\begin{equation}
\pmatrix{0&1\cr -1&0},
\end{equation}
where $0$ and $1$ are $N$ dimensional zero and unit matrices.

We shall use the classical geometric formulation of a quantum dynamical system in order to derive an equation for the quantum evolution
 constrained on a submanifold of $R^{2N}$ that corresponds to pure coarse-grained states.

Consider first the example of a pair of qubits. In this case the subspace of product states $|\psi^1>\otimes|\psi^2>$ is characterized
 by the following condition: $c_1c_4=c_2c_3$ on the expansion coefficients in the computational basis, which can be translated into a
 condition on the real canonical coordinates. This condition characterizes the submanifold of $R^{2N}$ of points corresponding to the
product states. The characterization of product, i.e. nonentangled, states is also given by the condition of maximality of $P_g(\psi)$ where
 $g$ is the algebra of local operators generated by $\sigma^1_{x,y,z}\otimes 1,\>1\otimes\sigma^2_{x,y,z}$.
 In terms of the canonical coordinates $(q,p)$ we obtained the purity function   $P_g(q,p)$.
 In fact local purities $<\sigma^{1,2}_x>^2+<\sigma^{1,2}_y>^2+<\sigma^{1,2}_z>^2$
 as functions of the canonical coordinates are equal and are
 represented by half of the total purity function.
It can be demonstrated that the function $P_g(p,q)$
is maximized precisely when the separability constraint $c_1c_4=c_2c_3$ is satisfied

Hamiltonian equations with the algebraic constraint
$c_1c_4=c_2c_3$ have been studied for the first time in
\cite{JaAnnPhys} and in \cite{Hughston_sym}. The formalism of
quantum constrained dynamics developed in \cite{JaAnnPhys} is
based on Dirac's approach to classical constrained Hamiltonian
systems and requires the constraint to be given explicitly in terms of
 an even number of independent real
functions. In our present case there is in general only one real
constraint: $\Phi(x)\equiv P_g(x)-max=0$, and the approach with
symplectic constraints of \cite{JaAnnPhys}, \cite{Hughston_sym}
 can not be applied. However, the formalism of the so called metrical constraints, developed in \cite{Hughston_met},
 with an arbitrary number of real functions defining the constraint submanifold is applicable.  For the benefit of the reader
we shall rederive the constrained dynamical equations with only one real constraint which is of interest here.

The purity constraint
 \begin{equation}
\Phi(x)=P_g(x)-max=0
 \end{equation}
represents a single scalar condition that we want to impose on the evolution. In order to impose this condition
 the component of the Hamiltonian vector field
$\dot x$ (4) normal to the constraint submanifold has to be removed resulting in
\begin{equation}
\dot x^i=\omega^{ij}\nabla_j {\cal H}-\lambda g^{ij}\nabla_j \Phi,
\end{equation}
where $g^{ij}$ is the unit metric on $R^{2N}$ and $\lambda$ is a single Lagrange multiplier to be determined.
Substitution of (7) in $\dot \Phi(x(t))$ results in
\begin{equation}
 \omega^{ij}\nabla_i\Phi\nabla_j {\cal H}=\lambda g^{ij}\nabla_i\Phi\nabla_j \Phi,
\end{equation}
from which
\begin{equation}
 \lambda={\omega^{ij}\nabla_i\Phi\nabla_j {\cal H}\over g^{ij}\nabla_i\Phi\nabla_j \Phi}.
\end{equation}
Substituting this $\lambda$ in (7) results in the constrained dynamical equations
\begin{equation}
 \dot x^i=\omega^{ij}\nabla_j {\cal H}-
{\omega^{ij}\nabla_i\Phi\nabla_j {\cal H}\over g^{ij}\nabla_i\Phi\nabla_j\Phi} g^{ij}\nabla_j \Phi.
\end{equation}

 We propose the reduction of the constrained equation (10) on the constrained manifold to represents dynamical
 equation of the coarse-grained pure states.

Observe that the numerator in (10) represent the Poison bracket $\{\Phi,{\cal H}\}=\dot \Phi$ and the denominator is
 $||\nabla \Phi||^2$. Using the equalities
\begin{equation}
L_{ij}q_j=\delta_{ij}{\partial <L>\over \partial q_j},\quad L_{ij}p_j=\delta_{ij}{\partial <L>\over \partial p_j},
\end{equation}
where $L_{ij}$ are matrix elements of the operator $L$
and in our case $\Phi(\psi)=P(\psi)-max$ the denominator can be further transformed as follows
\begin {eqnarray}
 &g^{ij}& \sum_{l,k}\nabla_i<L_l>^2\nabla_j<L_k>^2\nonumber\\
&=&4\sum_{l,k}<L_lL_k>-<L_l><L_k>=4\sum_l(\Delta L_l)^2=4\Delta(\psi).
\end {eqnarray}

Before presenting few examples we would like to make some comments concerning the constrained equation (10).

$1^o$ In the open system picture of the distinguished system, to be discussed in the next section, and in the
 usual weak coupling approximation (WCA), with the distinguished observables identified with the Lindblad generators,
 the above equation is greatly simplified.
Namely, in the WCA the Hamiltonian and the system operators $L_l$ that couple with the environment operators satisfy
\begin{equation}
[H,L_l]=\lambda_lL_l
\end{equation}
and  $\dot \Phi$ satisfies
\begin{equation}
 \dot \Phi=\dot P(\psi) =2\sum (\Delta L_l)^2=2\Delta(\psi),
\end{equation}
so that the equation (10) is reduced to
\begin{equation}
 \dot x^i=\omega^{ij}\nabla_j {\cal H}-
{1\over 2} g^{ij}\nabla_j \Phi.
\end{equation}
The open system interpretation of the constrained equation (10) and the equation (15) will be discussed in more
 details later in the next section.

$2^o$ An equivalent constrained equations, of the same form as (10) and in the special case (15),
 are obtained if instead of the purity constraint $P(\psi)=max$ the
 constraint $\Delta(\psi)=\sum_l(\Delta L_l)^2=min$ is used. In particular, the special case equation (15), valid under the same
 conditions, with the use of (11) can be written in the form
\begin{equation}
{d |\psi>\over dt}=-i[H,\psi]+\sum_l( L_l^2+<L_l>^2-2<L_l>L) |\psi>.
\end{equation}

$3^o$ Number of variables and equations in (10), and in (15), can be reduced if, prior to imposing the constraints,
 the normalization of $<\psi|\psi>$ and
 the global phase invariance of $|\psi>$ are explicitly used. The Hamiltonian Schroedinger equation (4) is then formulated on $S^{2N-1}/S^1$ instead
 of $R^{2N}$. The constrained equations have the same form as in (10) with the appropriate symplectic $\omega^{ij}$ and metric $g^{ij}$ forms.
 An example is provided in the example b) below.

$4^o$ The geometric Hamiltonian formulation and the constrained equations can be generalized to an infinite dimensional Hilbert space.

Before we analyze an open system physical model of the coarse-grained dynamics let us present few examples of the g-constrained systems.

 {\it Examples}

a) The first example is trivial in the sense that all pure states are g-coherent,
 and serves the purpose of illustrating the self-consistency of the approach.
 Consider a single qubit with the Hilbert space $H^2$ and an arbitrary Hamiltonian $H$. As the algebra of distinguished
observables we take $g=su(2)$. The g-purity is $P_g(\psi)=<\sigma_x^2>+<\sigma_y^2>+<\sigma_z^2>$ and is maximal for any
pure state. In this case all pure states are g-nonentangled and g-coherent.

The constraint $\Phi=P_g(\psi)-max=0$ in the real canonical coordinates assumes the following form
\begin{equation}
(p_1^2+p_2^2+q_1^2+q_2^2)^2=2.
\end {equation}
 The gradient of the constrains is given by
\begin{eqnarray}
\nabla_{q_1}\Phi&=&q_1(p_1^2+p_2^2+q_1^2+q_2^2)\nonumber\\
\nabla_{q_2}\Phi&=&q_1(p_1^2+p_2^2+q_1^2+q_2^2)\nonumber\\
\nabla_{p_1}\Phi&=&q_1(p_1^2+p_2^2+q_1^2+q_2^2)\nonumber\\
\nabla_{p_2}\Phi&=&p_2(p_1^2+p_2^2+q_1^2+q_2^2)
\end{eqnarray}
and the Poisson bracket $\{\Phi,{\cal H}\}$ is zero for arbitrary Hamiltonian ${\cal H}=<H>$. Thus, the constraints are trivially satisfied, and the
 constraint dynamics is reduced to the linear Schroedinger part: $\dot x^i=\omega^{ij}\nabla_j {\cal H}$. This example extends to the general case
 of $H^N$ with the distinguished algebra $g=u(N)$.

b) As the second example we consider the system of two qubits with the distinguished algebra of local observables $g=su(2)\otimes su(2)$. In this
 case  g-entanglement is the standard bipartite entanglement. g-nonentangled are the product states. Subsequent formulas are simplified if the
 condition $<\psi|\psi>=1$ and the phase invariance are explicitly used.
 With this the system is reduced on the projective space $S^7/S^1$.
Purity
$P(\psi)=<(\sigma^1_x)>^2+<(\sigma^1_y)>^2+<(\sigma^1_z)>^2 +<(\sigma^2_x)>^2+<(\sigma^2_y)>^2+<(\sigma^2_z)>^2$ in the computational basis
 and in canonical coordinates $\{q_1,q_2,q_3,p_1,p_2,p_3\}$ of $S^7/S^1$ is represented by
\begin{eqnarray}
P(q,p)&=&1+4(2{\sqrt 2} p_1(p_2q_3-p_3q_2)+2q_3^2+p_1^2(p_2^2+q_2^2))\nonumber\\
&+&4(2{\sqrt 2} q_1(p_2p_3+q_2q_3)-q_1^2(p_2^2+q_2^2)-2p_3^2)
\end{eqnarray}
 $P(\psi)=max$ is equivalent to $c_1c_4=c_2c_3$ where $c_1,c_2,c_3,c_4$ are coefficients of $|\psi>$ in the computational basis, and the equation
of this constraint is equivalent to two real equations
\begin{equation}
{\sqrt 2}p_3=p_2q_1+p_1q_2,\qquad {\sqrt 2} q_3=q_1q_2-p_1p_2.
\end{equation}

As for the Hamiltonian we consider two typical examples
\begin{eqnarray}
H_s&=&\sigma^1_z+\sigma^2_z+\mu\sigma^1_z\sigma^2_z\\
H_{ns}&=&\sigma^1_z+\sigma^2_z+\mu\sigma^1_x\sigma^2_x\\
\end{eqnarray}

The reduced g-constrained dynamics in $\{q_1,q_2,p_1,p_2\}$ coordinates
 is equivalently described by the equations (10) and the metrical constraints (19),
 or by the symplectic constrained equations
 with the constraints (20). The constrained equations with constraints (20) turn out to be of
  a simpler form and are reproduced here. The details of the derivation have been presented in
  \cite{JaAnnPhys}.

For the Hamiltonian  $H_s$ the constrained equations read
\begin{eqnarray}
 \dot q_1&=&-{4\mu p_1q_1q_2+2\omega p_1[2+(p_2)^2+(q_2)^2]\over 2+ (p_2)^2+(q_2)^2},\nonumber\\
  \dot q_2&=&-{4\mu p_2q_1q_2-2\omega p_2[2+(p_1)^2+(q_1)^2]\over 2+ (p_1)^2+(q_1)^2},\nonumber\\
   \dot p_1&=&{2\mu q_2 [(q_1)^2-(p_1)^2-2]+2\omega q_1[2+(p_2)^2+(q_2)^2]\over 2+ (p_2)^2+(q_2)^2},\nonumber\\
   \dot p_2&=&{2\mu q_1 [(q_2)^2-(p_2)^2-2]+2\omega q_2[2+(p_1)^2+(q_1)^2]\over 2+
(p_1)^2+(q_1)^2}.
 \end{eqnarray}
and for the Hamiltonian $H_{ns}$.
\begin{eqnarray}
\dot q_1&=&{2\mu p_1[(p_2)^2+(q_2)^2-2)]-2\omega p_1[2+(p_2)^2+(q_2)^2]\over 2+ (p_2)^2+(q_2)^2},\nonumber\\
  \dot q_2&=&{2\mu p_2[(p_1)^2+(q_1)^2-2]-2\omega p_2[(2+(p_1)^2+(q_1)^2]\over 2+ (p_1)^2+(q_1)^2},\nonumber\\
   \dot p_1&=&{-2\mu q_1 [(q_2)^2+(p_2)^2-2]+2\omega q_1[2+(p_2)^2+(q_2)^2]\over 2+ (p_2)^2+(q_2)^2},\nonumber\\
   \dot p_2&=&{-2\mu q_2 [(q_1)^2+(p_1)^2-2]+2\omega q_2[2+(p_1)^2+(q_1)^2]\over 2+ (p_1)^2+(q_1)^2}.
 \end{eqnarray}
 There are also the equations expressing $\dot q_3$ and $\dot p_3$
 in terms of $q_1,q_2,p_1,p_2$, but the solutions of these are
  already given by the constraints.

The dynamics generated by (24) and (25) is illustrated in figures 1 and 2. In fig. 1 we illustrate the time series $q_1(t)$ for single typical orbit
 of (24) (fig.1a) and of (25)(fig.1 b).
In figures 2a,b,c,d the Poincare sections $q_2=0,p_2>0$ for $H_{ns}$ (25) are shown.
It should be observed that g-constrained dynamics of the symmetric Hamiltonian $H_s$ is regular, while that of the Hamiltonian $H_{ns}$ with no
 such symmetry displays typical properties of the Hamiltonian chaos. Thus, although the linear Schroedinger equation always generates an integrable
 Hamiltonian system, the coarse-grained quantum system evolving according to the constrained equations can display all complexities of typical
 chaotic dynamics.

c) In this example we again consider a system with $g=su(2)$ distinguished algebra but with the spin $s=1$ i.e. with $H^3$ Hilbert space.
As for the Hamiltonian we take a nonlinear expression of $su(2)$ generators
\begin{equation}
H=J_z-2J_x+\mu J_z^2
\end{equation}
When $\mu\neq 0$ the Schroedinger evolution with the Hamiltonian (26) does not preserve the $su(2)$-coherent states.
The set of $su(2)$-coherent states is preserved when $\mu=0$.

The g-constraint $\Phi(\psi)=P_{su(2)}(\psi)-1=<J_x^2>+<J_y^2>+<J_z^2>-1$ in the eigenbases of $J_z$ and in the real canonical coordinates of $R^6$ assumes the form
\begin{eqnarray}
4P_{su(2)}(q,p)&=&-4+p_1^4+p_3^4-2p_3^2q_1^2+q_1^4+8p_2p_3q_1q_2+2p_3^2q_2^2\nonumber\\
&+&2q_1^2q_2^2+2p_2^2(p_3^2+(q_1-q_3)^2)+4q_1q_2^2q_3+2(p_3^2-q_1^2+q_2^2)q_3^2\nonumber\\
&+&q_3^4+4p_1(p_2^2p_3-p_3q_2^2+2p_2q_2q_3)\nonumber\\
&+&2p_1^2(p_2^2-p3^2+q_1^2+q_2^2-q_3^2)
\end{eqnarray}

The Poison bracket of the constraint and the Hamiltonian, that is needed for the constraint equations (10), reads
\begin{equation}
\omega^{ij}\nabla_i\Phi\nabla_j {\cal H}=2\mu[(p_3q_1+p_1q_3)(q_2^2-p_2^2)+2p_2q_2(p_1p_3-q_1q_3)].
\end{equation}
We see that, when $\mu=0$, the Poison bracket (28) is zero and the g-constrained equations reduce to the Schroedinger equation. The squared norm of
the $\Phi$ gradient is given by somewhat complicated function of the canonical coordinates $(q,p)$ and will not be reproduced here.
We illustrate the form of simplified constrained equations (15) by the formula for $\dot p_1$
\begin{eqnarray}
\dot p_1&=&-{\sqrt 2}[(1+\mu)q_1-q_2]\nonumber\\
&+&p_1(p_2^2-p_3^2+q_1^2+q_2^2-q_3^2)-2p_2q_2q_3+p_3q_2^2-p_2^2p_3-p_1^3,
\end{eqnarray}
where the first line is the Hamiltonian term and the second line is from the gradient of the constraint.

We shall come back to this example in the next section.

\section{Open system model}

The coarse-grained system specified by the distinguished variables can be considered as an open system with the larger closed system
characterized by the full algebra $u(N)$. In the case when the Hilbert space can be split into the tensor product with one component
 corresponding to the distinguished reduced system the standard open system model of decoherence applies. This theory singles out a distinguished
 set of states, the pointer or the robust states, and characterizes them as pure states of the reduced open system that remain pure under evolution,
 or as states in which the reduced open system is not and does not get entangled with the environment during the full system evolution.
Reduced states of the general coarse-grained system, discussed in sections 2 and 3, satisfy the same properties as the robust states
of an open system under decoherence if the interaction of the open system and the environment is mediated by all of the distinguished observables. 
It has been demonstrated that the robust states in this case coincide
 with the g-coherent states\cite{Viola4} in the weak coupling limit.
In this picture the coarse-graining physically occurs due to decoherence of the distinguished system induced by specific interaction with the environment
 which represents generalized simultaneous measurement of all distinguished observables.
The pointer states are identified with reduction of the
 g-nonentangled or g-coherent states.

We would like to identify the distinguished observables with observables that are simultaneously measured on the open system. In the weak coupling limit (WCL) the Born-Markov and rotating wave approximations result in
 the Lindblad master equation of the open system dynamics \cite{Brauer}
\begin{equation}
 \dot \rho(t)=-i[H,\rho]+{1\over 2}\sum_l \left ([L_l\rho,L^{\dag}_l]+[L_l,\rho L^{\dag}_l\right),
\end{equation}
where $H$ is the open system Hamiltonian and $L_l$ are the so called Lindblad operators. $L_l$ are the
 open system operators that are coupled with that what is considered as environment. If the eq (30)
 corresponds to the measurement of certain observables than $L_l$ are the Hermitian operators that represent
 the measured observables. In our model of the coarse-graining we shall suppose that the distinguished
 algebra is precisely the algebra formed by the Hermitian Lindblad operators in (30).

 As pointed out, the pointer or robust states in the open system model of decoherence are the pure states
 of the open system that remain pure in the course of evolution.  It has been suggested \cite {Strunz}
 that an approximate
 evolution equation of the pure robust states can be obtained by minimizing the Hilbert-Schmidt (HS)
 distance from $\rho(t)$ to the set of pure states. In the case $\rho(t)$ is given by the Lindblad eq.
 with Hermitian Lindblad operators
 the equation  of the HS closest pure state is \cite{Strunz}
\begin{equation}
 {d|\psi>\over dt}=-i[H,\psi]+\sum_l( L_l^2+<L_l>^2-2<L_l>L_l) |\psi>.
\end{equation}
 This is precisely our constrained equation (15) when g-entanglement measure $P(\psi)=\sum_l <L_l>^2$ is
 replaced by the equivalent measure $\Delta(\psi)=\sum_l(\Delta L_l)^2$ in the WCL with the distingushed
 observables being the  Lindblad generators.

 The equation (31) (or (15)) represent the deterministic part of the
 stochastic Schroedinger equation derived in the quantum state diffusion theory \cite{QSD} for arbitrary
 random pure state, which we reproduce here because it will be used for numerical computations in the next
 example. The Ito form of the QSD equation corresponding to (30) reads
\begin{eqnarray}
|d\psi>&=&-i H |\psi>dt\nonumber\\
&+&\left [\sum_l 2< L_l^{\dag
}>  L_l- L_l^{\dag} L_l
-< L_l^{\dag}> < L_l>\right ]|\psi(t)>dt\nonumber\\
&+& \sum_l ( L_l-<L_l>)|\psi(t)> dW_l
\end{eqnarray}
where $dW_l$ are independent increments (indexed by $l$)
of complex Wiener c-number processes $W_l(t)$ satisfying
\begin{eqnarray}
{\rm E}[dW_l]={\rm E}[dW_ldW_{l'}]&=&0,\nonumber\\
dW_l{ d\bar W}_{l'}&=& \delta_{l,l'}dt,\nonumber\\
l&=&1,2\dots m,
\end{eqnarray}
where $E[\cdot]$ denotes the expectation with respect to the
probability distribution given by the (m-dimensional)  process
$W$,  and  $\bar W_l$ is the complex conjugate of $ W_l$.

The random vector $|\psi(t)>$ which satisfies (32) is related to the density matrix
 $\rho(t)$ which satisfies
 the Lindblad equation (30) by averaging over the realizations of the process (32)
\begin{equation}
 \rho(t)=E[|\psi(t)><\psi(t)|].
\end{equation}

Let us stress that the HS approximate robust state eq. (31) assumes validity of WCL and coincides with the
 constrained eq. (15) simplified from (10) under this assumption. On the other hand the general constrained evolution given by
 (10) is valid, in the sense that it preserves $P(\psi)$ and $\Delta(\psi)$,
 with no assumption about special evolution of $\dot P(\psi)$ which is obtained under the WCL.

 {\it An example}

In the case of an open system that satisfies the conditions for the weak coupling approximation (13) the dynamics of the system
 is described well by the Lindblad equation and the pointer states are exactly the g-coherent states \cite{Viola4}. 
Using particular examples, it has been demonstrated \cite{Strunz}
 that the equation (31), which coincides with the simplified form of the constrained equation (15), describes
well the evolution of the pointer i.e. g-coherent states. We shall analyze here an example that does not satisfy the condition (13)
 of the WCA.

Let us consider, as an example, the two mode Bose-Hubbard model (see for example \cite{BH}), given by the following Hamiltonian with
$\hbar=1$
\begin{equation}
H=\epsilon_1a_1^{\dag}a_1+\epsilon_2a_2^{\dag}a_2+\alpha (a_1^{\dag}a_2+a_2^{\dag}a_1)+\mu(a_1^{\dag
2}a_1^2+a_2^{\dag 2}a_2^2),
\end{equation}
where $a_i,a_i^{\dag},i=1,2$ are bosonic annihilation and creation operators of the two modes. The dynamics
 preserves total particle number $N=a_1^{\dag}a_1+a_2^{\dag}a_2$. Introducing operators
\begin{equation}
 q_j=(a_j^{\dag}+a_j)/\sqrt 2,\>
p_j=i(a_j^{\dag}-a_j)/\sqrt 2 ,\> j=1,2,
\end{equation}
or the operators
\begin{eqnarray}
J_x&=&{1\over 2}(a_1^{\dag} a_2+a_2^{\dag} a_1)\nonumber\\
J_y&=&{i\over 2}(a_1^{\dag} a_2-a_2^{\dag} a_1)\nonumber\\
J_z&=&{1\over 2}(a_2^{\dag} a_2-a_1^{\dag} a_1)),
\end{eqnarray}
the Hamiltonian assumes the following forms respectively in coordinates (36)
\begin{eqnarray}
H&=&\epsilon_1(p_1^2+q_1^2)/2+\mu(p_1^2+q_1^2)^2/4+\nonumber\\
&+&\epsilon_2(p_2^2+q_2^2)/2+\mu(p_2^2+q_2^2)^2/4+\nonumber\\
&+&\alpha (p_1p_2+q_1 q_2)
\end{eqnarray}
and in terms of (37)
\begin{equation}
H=-2\alpha
J_x+2(\epsilon_2-\epsilon_1)J_z+\mu J_z^2.
\end{equation}
In what follows we shall always set $\alpha=1,\epsilon_2-\epsilon_1=1$.

The preserved total number of particles is related to $J^2$ by $J^2=N/2(N/2+1)$. 
 Thus, the effective Hilbert space of the system carries an irreducible representation of $SU(2)$, which is the dynamical
group of the model. This suggest that the $SU(2)$ coherent states have a special status in the model (35).
This however is not true, because the nonlinear term $\mu J_z^2$ makes the set of $SU(2)$ coherent
 states noninvariant.

We would like to analyze system (35) interacting with an environment via operators (37) or (36). The Hamiltonian (35) and operators (36) or (37) used as the Lindblad operators
 do not quite satisfy the condition (13) for the WCA. Nevertheless, we shall suppose that the open system evolution is described by the Lindblad equation
 with Lindblad operators given either by (36) or by (37). Notice that the result $\Delta_g(\psi)\rightarrow min$ obtained in \cite{Viola4}
 does not apply necessarily since the system does not satisfy the WCA condition.
 We shall demonstrate that the asymptotic states of the Lindblad eq.
 of an open BH system interacting
 with an environment via the Lindblad operators $L_l$ satisfy the constraints condition $\Delta_g(\psi)=min$ almost exactly

Let us first consider the open system evolution in terms of random pure states $|\psi(t)>$ and the QSD equation (32).
 We first choose $L_1=J_x,L_2=J_y,L_3=J_z$ and compute $\Delta_{su(2)}(\psi(t))$ and
$P_{su(2)}(\psi(t))$ from an initial state equal to the number state given by
$(a_1^{\dag})^2(a_2^{\dag})^2|0,0>$. The results are shown in figure 3. The state quickly converges to
 those with a minimal $\Delta_{su(2)}(\psi(t))$, i.e. to the $su(2)$-coherent states.
 On the other hand $\Delta_{H_4}=\Delta^2p_1+\Delta^2p_2+\Delta^2q_1+\Delta^2q_2$
 remains constant and large.
$su(2)$-purity is less than maximal at the beginning but quickly converges to the maximal value. Although
 the state $|\psi(t)>$ is always a pure state of the Hilbert space, its $su(2)$-purity is maximal only when
 $|\psi>$ is an $su(2)$-coherent  i.e. an $su(2)$-nonentangled state.
Analogously,
assuming the Lindblad operators to be $L_1=q_1,L_2=q_2,L_3=p_1,L_4=p_2$ implies an evolution such that
 $\Delta_{su(2)}$ is far away from its minimum, but $\Delta_{H_4}$ converges to values close to the minimal
 and remains such for almost all times (please see fig. 4).

The equivalent conclusions are obtained when the evolution is described in terms of
$\rho(t)=E(|\psi(t)><\psi(t)|)$, i.e. by the Lindblad equation. This is illustrated in figure 5 and 6,
 with $\Delta_{g}$ for $g=su(2); g=H_4$ and  $L_1=J_x,L_2=J_y,L_3=J_z$ 
with the number initial state (fig. 5),
 and $L_1=q_1,L_2=q_2,L-3=p_1,L_4=p_2$ with an $su(2)$ coherent initial state in figure 6. 
Only two hundred QSD sample paths are use to compute $\rho(t)$ and then the corresponding $\Delta_g(\rho)$

Furthermore, consider evolution from an $su(2)$ coherent initial state with $J=1$ and with the Lindblads being
  $J_{x,y,z}$. The Lindblad eq. assumes WCL, the asymptotic states satisfy $\Delta_{su(2)}\approx min$,
  and the simplified constrained equation (15) applies.
 Indeed, the Lindblad evolution is well approximated by the simple form of the constrained equation (15), as is illustrated in figure 7a,b.

The usual picture of decoherence applies: An arbitrary initial state evolves very quickly into a
 mixture of $g$-coherent states, and then each of these evolves in a way that is well approximated by
 the nonlinear eq. (10) or in the WCL by (15).

Let us stress that the coarse-graining by distinguished observables, discussed here,
is specially appropriate in a description of macroscopic features of a quantum system,
 with the distinguished observables identified with the macroscopic quantities. In this case
 the Hilbert space of the quantum system does not have the bipartite tensor product structure,
 with one party being characterized by the macroscopic observables and the other party
being the environment. The usual models of decoherence \cite{Schloss} with the initial separation
$|\psi>=|\psi_s>\otimes |\psi_{env}>$ do not apply. However, the picture of coarse-graining by
 distinguished
 observables with the corresponding nonlinear evolution equations can be applied.

\section{Summary}

We have analyzed the coarse-graining introduced by a chosen set of distinguished observables. The
algebra of distinguished observables defines the corresponding generalized nonentangled states which
coincide by definition with the generalized coherent states. The states obtained by reduction on the
 distinguished observables of the g-nonentangled states  are pure. We have propose to consider the
 coarse-grained evolution as constrained Schroedinger dynamics, where the constraint guaranties
 that the state is always pure g-nonentangled. In order to formulate the constrained evolution equations
we used Hamiltonian formulation with the metrical form of the
 constrained dynamics as developed in \cite{Hughston_met}.

Further on we discussed an open system model of the coarse-graining and of the reduced constrained
 equation. In the weak coupling limit the open system dynamics is given by the Lindblad master equation.
In this limit, and if the Lindblad operators are taken to represent the distinguished observables then the
 constrained equations for the g-coherent states developed here coincide with previously suggested \cite{Strunz} evolution equation for
 the pointer states of the open system.

Our simplified constrained evolution equation (15) for the g-nonentangled states coincides with the deterministic part
 of the Ito stochastic Schroedinger equation developed in the quantum state diffusion theory (QSD) of open system
 dynamics.The stochastic Schroedinger equation describes dynamics of any random pure state. Our
 constrained equations describe dynamics of deterministic pure states which are in the subset of all
 pure states that remain pure during the evolution. From a formal point of view, it would be interesting to
 derive the QSD stochastic equations using the formalism of constraints, where the constraint would be given
 by random variables representing the obtained results of measurements with Gaussian distribution.

The formalism of coarse-graining as the constrained evolution can be used to study coarse-grained
 macroscopic observables of a quantum system and derive their classical behavior.
\vskip 1cm

 {\bf Acknowledgments} This work is
partly supported by the Serbian Ministry of Science contract No.
141003. I should also like to acknowledge the support and
hospitality of the Abdus Salam ICTP.

\newpage
{\bf FIGURE CAPTION}

Figure 1. Illustrated are segments of time series $q_1(t)$ for the system (24) in a) and
 (25) in b).

Figure 2. Poincare sections $q_2=0,p_2>0$ and ${\cal H}=1.5$ for the system (25).  The parameters are
 (a)$\mu=1.1$, (b) $\mu=1.3$, (c) $\mu=1.5$
and (d) $\mu=1.7$

Figure 3. Illustrates the invariant fluctuation $\Delta_g (\psi)$ (2) in the cases
$g=su(2)$ (full line) and $g=H_4$ (dotted line)  for the QSD evolution
 with the Hamiltonian (35) and Lindblads $L_1=J_x,L_2=J_y,L_3=J_z$.
The initial state is the number state $|2,2>=(a_1^{\dag})^2(a_2^{\dag})^2|0,0>$.
 The parameters are $\mu=0.1,\alpha=1,\epsilon=0,\gamma=0.9$

Figure 4. Illustrates the invariant fluctuation $\Delta_g (\psi)$ (2) in the cases
$g=su(2)$ (full line) and $g=H_4$ (dotted line)  for the QSD evolution
 with the Hamiltonian (35) and Lindblads $L_1=q_1,L_2=q_2,L_3=p_1,L_4=p_2$. The initial states is an $su(2)$ coherent state.
 The parameters are $\mu=0.1,\alpha=1,\epsilon=0,\gamma=0.9$

Figure 5. Illustrates the invariant fluctuation $\Delta_g (\rho)$ (2) in the cases
$g=su(2)$ (full line) and $g=H_4$ (dotted line)  for the evolution
 by the Lindblad eq. with the Hamiltonian (35) and Lindblads $L_1=J_x,L_2=J_y,L_3=J_z$.
The initial state is the number state $|2,2>=(a_1^{\dag})^2(a_2^{\dag})^2|0,0>$.
 The parameters are $\mu=0.1,\alpha=1,\epsilon=0,\gamma=0.9$.

Figure 6. Illustrates the invariant fluctuation $\Delta_g (\rho)$ (2) in the cases
$g=su(2)$ (full line) and $g=H_4$ (dotted line)  for the evolution
 by the Lindblad eq. with the Hamiltonian (35) and Lindblads $L_1=q_1,L_2=q_2,L_3=p_1,L_4=p_2$. 
The initial states is an $su(2)$ coherent state.
 The parameters are $\mu=0.1,\alpha=1,\epsilon=0,\gamma=0.9$

Figure 7. Evolution of $Tr[\rho\sigma_z]$ according to the Lindblad eq. (30) (full line) and
 of $<\sigma_z>$ according to the simplified constrained eq. (15) (dotted line) with the Hamiltonian (35)
 and Lindblads $L_1=J_x,L_2=J_y,L_3=J_z$. The initial states is $su(2)$ coherent state $|j,j_z>=|1,-1>$. The parameters are
$\mu=0.1,\alpha=1,\gamma=0.2$ and a) $\epsilon=0$ and b) $\epsilon=1$.

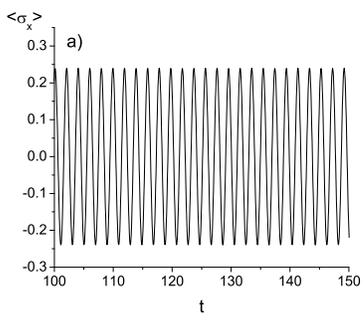
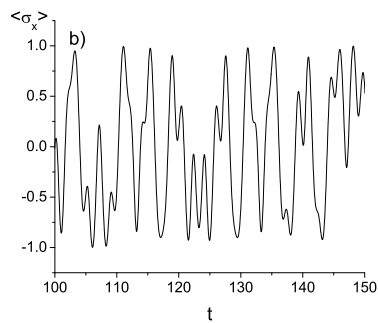
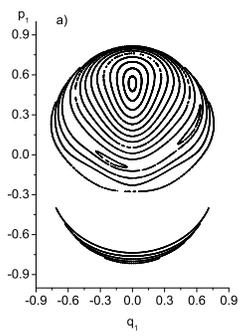
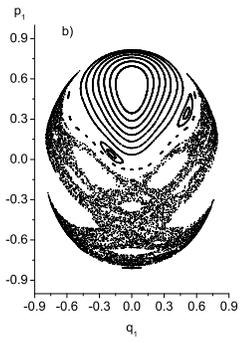
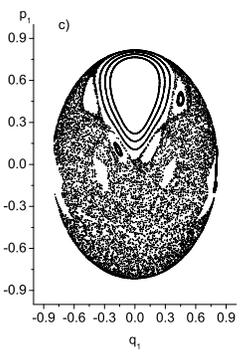
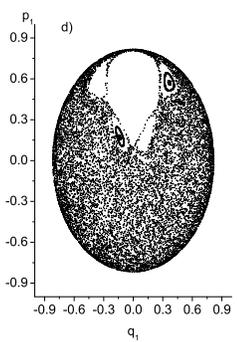
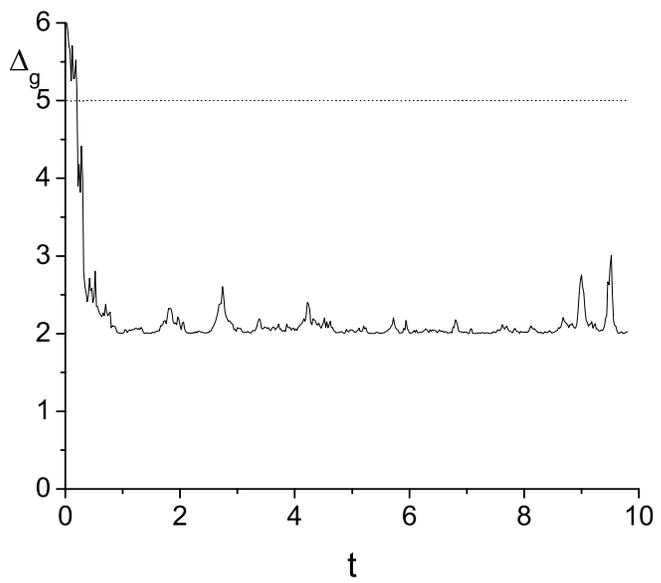
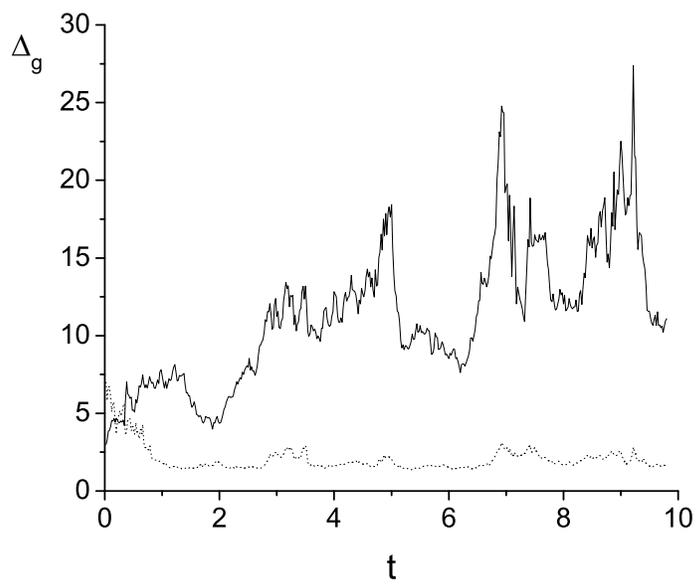
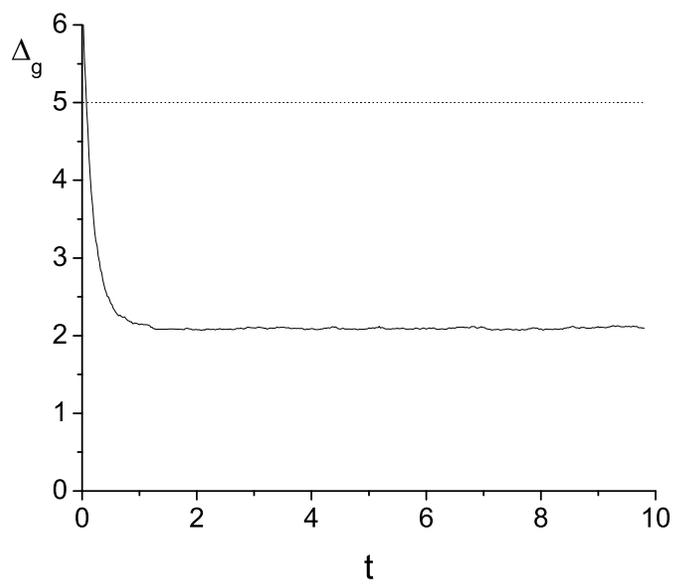
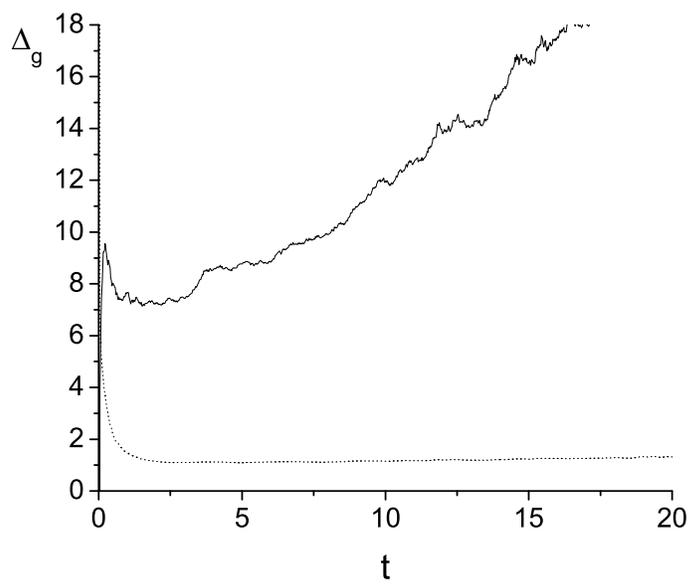
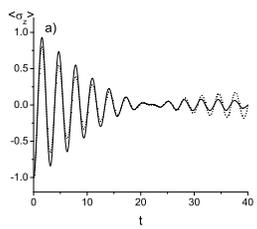
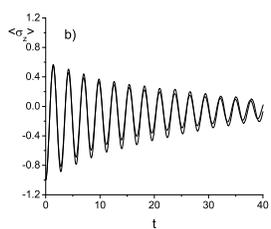
\end{document}